\documentclass[conference]{IEEEtran}
\IEEEoverridecommandlockouts
\usepackage{cite}
\usepackage{amsmath,amssymb,amsfonts}
\usepackage{algorithmic}
\usepackage{graphicx}
\usepackage{textcomp}
\usepackage{xcolor}
\usepackage{tablefootnote}
\usepackage{multirow}%
\usepackage{manyfoot}%
\usepackage{booktabs}%
\def\BibTeX{{\rm B\kern-.05em{\sc i\kern-.025em b}\kern-.08em
    T\kern-.1667em\lower.7ex\hbox{E}\kern-.125emX}}
\begin{document}

\title{Local Conformity-Based Evolutionary Game Modeling of UAV Swarm Under Byzantine Attack

\thanks{Corresponding author: Junhui Zhao.
	}
}

\author{
	\IEEEauthorblockN{
		Ruixing Ren\IEEEauthorrefmark{1}, 
		Junhui Zhao\IEEEauthorrefmark{1}, 
		and He Fang\IEEEauthorrefmark{2}} 
	\IEEEauthorblockA{\IEEEauthorrefmark{1}School of Electronic and Information Engineering, Beijing Jiaotong University, Beijing 100044, China.}
	\IEEEauthorblockA{\IEEEauthorrefmark{2}College of Computer and Cyber Security, Fujian Normal University, Fuzhou 350117, China.}
	\IEEEauthorblockA{Emails: renruixing0604@163.com, junhuizhao@hotmail.com, fanghe@fjnu.edu.cn}
}

\maketitle

\begin{abstract}
Leveraging their flexible and efficient deployment capabilities, unmanned aerial vehicle (UAV) swarms have been widely applied in various mission scenarios. However, the open communication environment also exposes them to the threat of Byzantine attacks. Most existing studies assume independent decision-making by each UAV, neglecting that local conformity amplifies false information propagation. This paper constructs an evolutionary game model for UAV swarm under malicious attacks based on graph evolutionary game theory, revealing how local conformity rules govern the spread of deceptive strategies. Using death-birth updating rules, we derive the macroscopic dynamic equation for the fraction of deceptive strategies and the analytical solutions to its evolutionary stable states. Simulations reveal observation errors weaken malicious induction, while higher proportions of malicious nodes and greater attack intensity drastically amplify attack impacts. Moreover, the model exhibits strong topological robustness across regular, scale-free and random networks.
\end{abstract}


\section{Introduction}
With distributed architecture, high mobility and powerful
coordination, unmanned aerial vehicle (UAV) swarm exhibits
great potential in missions including cooperative reconnais
sance and target tracking \cite{RenUAV}. However, open communication renders swarms highly susceptible to Byzantine attacks from malicious nodes \cite{Fanwei}: by injecting false data, adversaries
mislead legitimate UAVs away from their true states and
impair mission efficiency \cite{Wang,RenLow}.

Most conventional security mechanisms assume independent decision-making of each UAV and overlook local interactions induced by communication topologies in practical swarms. For instance, interactions among swarm UAVs in \cite{Hu} only involve leader election and authentication information forwarding, without persistent tightly coupled collaborative decision-making. In \cite{Qirui}, UAV decisions rely on global optimization objectives and prediction models; communication topologies serve information exchange, yet security decisions barely depend on real-time local neighbor games or consensus. Though clustering and communication link modeling are adopted in \cite{Yang}, security decisions are derived via centralized or block coordinate descent optimization. Each UAV acts independently within the given optimization framework, with no emphasis on dynamic coupling of local interactions.

In practical swarm control, local conformity rules (the tendency to align states with neighbors) are commonly embedded in algorithms to boost coordination efficiency \cite{Chenrui,Gong}, and widely adopted for formation maintenance and consensus achievement \cite{IOTM,Zhaoplatooning}. Nevertheless, malicious nodes may exploit this rule: after malicious UAVs transmit false reports, legitimate UAVs may follow suit and broadcast erroneous information under consensus effects, triggering information cascade propagation \cite{Huang}. Few existing works analyze this process from an evolutionary dynamics perspective\cite{Wu}.

Evolutionary game theory provides a natural framework to characterize individuals’ strategy updating driven by neighbors’ payoffs, and has been applied to analyze strategy diffusion and consensus formation over networks \cite{DB,Lin}. Built on this theory, this work constructs an evolutionary game model incorporating local conformity behaviors for UAV swarm with malicious attackers, and investigates the influences of such behavior on Byzantine attacks. The main contributions are summarized as follows:
\begin{itemize}
	\item We integrate local conformity rules with Byzantine attacks to establish an evolutionary game framework for interactions between legitimate UAVs and malicious nodes, uncovering the propagation mechanism of deceptive strategies within swarms.
	\item Based on death-birth updating rules, we derive linear differential equations governing the temporal evolution of the fraction of deceptive strategies among legitimate UAVs, and obtain analytical expressions for evolutionary stable states (ESS).
	\item Monte Carlo simulations indicate observation error, malicious node fraction and attack intensity dominate steady-state false alarm rates, and the model sustains strong stability across diverse network topologies.
\end{itemize}

The remainder of this paper is organized as follows. Section \ref{Sec2} presents the system model and graph evolutionary game formulation. Section \ref{Sec3} derives the evolutionary dynamics equations and corresponding ESS for UAV swarms. Section \ref{Sec4} validates the theoretical model via simulations. Section \ref{Sec5} concludes this work.

\section{Graph Evolutionary Game Formulation}\label{Sec2}
Consider a swarm network $\mathcal{V}$ consisting of $N$ UAVs. Each UAV measures the binary system state $\theta^t\in\{0,1\}$ at each discrete time step $t$ and reports measurements to the ground station, where $\{\theta^t\}$ follows an independent and identically distributed process. Sensor noise, positioning bias and electromagnetic interference induce observation errors for each UAV, with the error probability denoted as $\varepsilon$.

The measurement of UAV $i\in\mathcal{V}$ at time $t$ is $u_i^t$, satisfying $\mathbb{P}(u_i^t\ne\theta^t)=\varepsilon$. After measurement, each UAV transmits a report $r_i^t$ to the ground station. Malicious UAVs execute attack strategies, while legitimate UAVs follow local conformity rules whose strategies evolve dynamically under Byzantine attacks launched by malicious UAVs. The ground station aggregates all reports for data fusion to infer the true system state. Evolutionary game theory characterizes incremental strategy updates via local interactions, matching the neighbor-aware distributed decision scheme of UAV swarms. Accordingly, this work adopts graph evolutionary game theory to model interactions between legitimate UAVs and their neighbors, including malicious nodes.

\textbf{Strategy Space.} Legitimate UAVs have two available strategies. The normal strategy $S_n$ reports raw measurements directly, namely $r_i^t=u_i^t$. The deceptive strategy $S_m$ sends inverted readings such that $r_i^t=\bar{u}_i^t$. This deceptive strategy does not stem from active malice of legitimate UAVs; instead, it represents erroneous decisions induced by neighboring agents, particularly malicious UAVs, under local conformity rules. Malicious UAVs adopt attack strategy $S_a$ and deliberately flip their reported measurements with probability $P_a$. Let $p_m$ denote the fraction of legitimate UAVs that employ the deceptive strategy, and $\beta$ represent the ratio of malicious UAV count to legitimate UAV count.

\textbf{Mission Payoff.} When UAVs interact with their neighbors, they receive mission payoffs based on whether their reports align with those of adjacent agents. Consider a focal UAV adopting the normal strategy $S_n$: it gains payoff $u_{ns}$ if its report matches neighbors’ reports from the previous time step, and $u_{nd}$ otherwise. If the focal UAV selects the deceptive strategy $S_m$, it obtains $u_{ms}$ for consistent neighbor reports and $u_{md}$ for inconsistent ones. To fulfill collaborative mission demands, legitimate UAVs are inherently designed to favor alignment with neighbors for consensus achievement, which yields the inequalities $u_{ns}>u_{ms}$ and $u_{nd}>u_{md}$. Furthermore, consistent reporting generates higher payoffs than mismatched reports, such that $u_{ns}>u_{nd}$ and $u_{ms}>u_{md}$.

\textbf{Utility Function.} Within the framework of evolutionary game theory, the utility function quantifies each player’s payoff level. Strategies yielding higher utilities enjoy evolutionary advantages and are more likely to be adopted across the population. The utility function is defined as follows:
\begin{equation}
	\pi=(1-\alpha)B+\alpha\cdot U
\end{equation}
where $B$ denotes the baseline fitness reflecting the inherent attributes of each player. A player that is less susceptible to the influence of other agents possesses a relatively large baseline fitness $B$. Considering network homogeneity, this work assumes identical baseline fitness for all UAVs, namely $B=1$. The parameter $\alpha$ stands for the weak selection coefficient, which quantifies the consensus tendency of UAVs and generally takes a very small value. $U$ measures the total payoffs acquired from interactions with neighboring UAVs.

UAVs cannot acquire full knowledge of their neighbors’ strategy profiles and only observe neighbors’ reported measurements. Therefore, each UAV evaluates its neighbors’ utilities based on its own sensor readings: neighbors whose reports match its local measurements are regarded as adopting the honest strategy $S_n$, while those with mismatched reports are deemed to employ the deceptive strategy $S_m$. This setting complies with the limited information constraint of distributed sensing systems.

\textbf{Strategy Updating Rule.} During long-term iterative evolution, legitimate UAVs continuously adjust their decision-making strategies under the influence of neighboring nodes. Combined with the inherent characteristics of UAV swarm scenarios, this paper adopts the death-birth updating rule to characterize the strategy evolution of each node, which is detailed as follows:
\begin{itemize}
	\item death step: In each iteration, a legitimate UAV is randomly selected to abandon its current behavioral strategy temporarily;
	\item birth step: The selected UAV imitates the strategy of a neighboring node and updates its own decision with a probability proportional to the neighbor’s fitness.
\end{itemize}

\textbf{ESS.} It denotes a stable equilibrium reached by the system in the evolutionary process. Once the system converges to an ESS, the system can automatically counteract perturbations and revert to its steady state even if a small fraction of nodes produce behavioral mutations. In this work, the system arrives at an ESS when the fraction $p_m^t$ of legitimate UAVs adopting deceptive strategies remains constant over time, which satisfies the evolutionary dynamics equation below:
\begin{equation}
	\left. \hat{p}_m^t \right|_{p_m^t = p_m^{\text{ESS}}} = 0
\end{equation}
where $\hat{p}_m^t$ represents the time derivative of $p_m^t$.

\section{Evolutionary Dynamics and Stable States of UAV Swarm}\label{Sec3}
Built upon the above model setup, this section analyzes how legitimate UAVs gradually adopt erroneous deceptive strategies under the misguidance of malicious UAVs, and derives the corresponding final stable states.

\subsection{State Transition Probability}
Suppose among the $k$ neighbors connected to a focal UAV, $k_m$ neighbors adopt strategy $S_m$ and alter their output readings, while the remaining $k_n=k-k_m$ neighbors employ strategy $S_n$.
Given the observation error probability $\varepsilon$, let $k^S_n$ denote the number of neighbors whose reported measurements coincide with the true system state, and $k^S_m=k-k^S_n$ denote the count of neighbors with reports inconsistent with the true system state. We thus obtain
\begin{equation}
	k^S_m=(1-\varepsilon)k_m+\varepsilon k_n,
\end{equation}
\begin{equation}
	k^S_n=(1-\varepsilon)k_n+\varepsilon k_m.
\end{equation}

In practical, UAVs cannot judge whether their observations contain measurement errors. Accordingly, we split the utility function into two cases for analysis: adopting strategy $S_n$ or $S_m$ under error-free observations and erroneous observations. When the focal UAV obtains error-free measurements and selects the deceptive strategy $S_m$, we have $r_i^t=\bar{u}_i^t\ne\theta^t$, and the corresponding utility function is written as
\begin{equation}\label{5}
	\pi_m^{(A)}=1-\alpha+\alpha\left [ k^S_m u_{ms}+ (k-k^S_m)u_{md}\right ].
\end{equation}
If the focal UAV adopts the normal strategy $S_n$, namely $r_i^t=u_i^t\ne\theta^t$, its utility function reads
\begin{equation}\label{6}
	\pi_n^{(A)}=1-\alpha+\alpha\left [ k^S_n u_{ns}+ (k-k^S_n)u_{nd}\right ].
\end{equation}
Similarly, when the focal UAV suffers from observation errors, the corresponding utility functions are given by
\begin{equation}\label{7}
	\pi_m^{(B)}=1-\alpha+\alpha\left [ k^S_n u_{ms}+ (k-k^S_n)u_{md}\right ],
\end{equation}
\begin{equation}\label{8}
	\pi_n^{(B)}=1-\alpha+\alpha\left [ k^S_m u_{ns}+ (k-k^S_m)u_{nd}\right ].
\end{equation}

Following the birth-death updating rule, we define $P_{n\to m}$ as the probability that a UAV switches from the normal strategy to the deceptive strategy. The numerator equals the sum of utilities of neighbors judged to adopt strategies conflicting with the true system state, while the denominator, denoted as $\Omega$, is the total utility of all neighboring nodes. When the focal UAV has no observation error, this probability can be expressed as
\begin{equation}\label{9}
	P^{(A)}_{n\to m}=\frac{k_m(1-\varepsilon)\pi^{(A)}_m+k_n\varepsilon\pi_n^{(B)}}{\Omega},
\end{equation}
\begin{equation}\label{10}
	\begin{aligned}
		\Omega &=\left [ k_n(1-\varepsilon)\pi^{(A)}_n+k_m\varepsilon\pi_m^{(B)} \right ]\\
		&+\left [ k_m(1-\varepsilon)\pi^{(A)}_m+k_n\varepsilon\pi_n^{(B)} \right ].
	\end{aligned}
\end{equation}
When the focal UAV suffers observation errors, the probability takes the form
\begin{equation}\label{11}
	P^{(B)}_{n\to m}=\frac{k_n(1-\varepsilon)\pi^{(A)}_n+k_m\varepsilon\pi_m^{(B)}}{\Omega}.
\end{equation}
Combining Eqs. (\ref{9}) and (\ref{11}), we can derive the probability expression for any legitimate UAV in the network switching from the normal strategy to the deceptive strategy as follows:
\begin{equation}\label{12}
	P_{n\to m}=(1-\varepsilon)P^{(A)}_{n\to m}+\varepsilon P^{(B)}_{n \to m}
\end{equation}
To further analyze Eq. (\ref{12}), we substitute Eqs. (\ref{5})-(\ref{8}) into Eqs. (\ref{9})-(\ref{11}), then plug the resulting expressions into Eq. (\ref{12}). After algebraic simplification, we obtain:
\begin{equation}\label{13}
	P_{n \to m} = \frac{
		\bar{\varepsilon}^2 k_m \pi_n^{(A)} + \varepsilon^2 k_m \pi_m^{(B)} + \varepsilon \bar{\varepsilon} k_n \pi_n^{(B)} + \varepsilon \bar{\varepsilon} k_n \pi_m^{(A)}
	}{
		\bar{\varepsilon} k_n \pi_n^{(A)} + \varepsilon k_n \pi_n^{(B)} + \bar{\varepsilon} k_m \pi_m^{(A)} + \varepsilon k_m \pi_m^{(B)}
	},
\end{equation}
where $\bar{\varepsilon}=1-\varepsilon$. Notice that Equations (\ref{5})-(\ref{8}) all contain a common term $1-\alpha$, followed by terms related to payoffs, which are denoted as $\Psi_i^{(X)}$. Furthermore, these expressions can be uniformly rewritten as
\begin{equation}\label{14}
	\pi^{(X)}_i=(1-\alpha)+\alpha\cdot\Psi^{(X)}_i,
\end{equation}
Substituting this unified expression into Eq. (\ref{13}), we obtain the numerator as
\begin{equation}
	\begin{aligned}
		&= \bar{\varepsilon}^2 k_m \left[ (1 - \alpha) + \alpha \Psi_n^{(A)} \right]
		+ \varepsilon^2 k_m \left[ (1 - \alpha) + \alpha \Psi_m^{(B)} \right]\\
		&+ \varepsilon \bar{\varepsilon} k_n \left[ (1 - \alpha) + \alpha \Psi_n^{(B)} \right]
		+ \varepsilon \bar{\varepsilon} k_n \left[ (1 - \alpha) + \alpha \Psi_m^{(A)} \right].
	\end{aligned}
\end{equation}
Meanwhile, we separate the terms containing $1-\alpha$ from those with $\alpha$:
\begin{equation}
	\begin{aligned}
		&= (1 - \alpha) \left( \bar{\varepsilon}^2 k_m + \varepsilon^2 k_m + \varepsilon \bar{\varepsilon} k_n + \varepsilon \bar{\varepsilon} k_n \right)\\
		&+ \alpha \left( \bar{\varepsilon}^2 k_m \Psi_n^{(A)} + \varepsilon^2 k_m \Psi_m^{(B)} + \varepsilon \bar{\varepsilon} k_n \Psi_n^{(B)} + \varepsilon \bar{\varepsilon} k_n \Psi_m^{(A)} \right).
	\end{aligned}
\end{equation}
After combining like terms, we arrive at
\begin{equation}
	= (1 - \alpha) \left( \bar{\varepsilon}^2 k_m + \varepsilon^2 k_m + 2\varepsilon\bar{\varepsilon} k_n \right) + \alpha \cdot \Phi_{1,}
\end{equation}
where,
\begin{equation}
	\Phi_{1} = \bar{\varepsilon}^2 k_m \Psi_n^{(A)} + \varepsilon^2 k_m \Psi_m^{(B)} + \varepsilon\bar{\varepsilon} k_n \Psi_n^{(B)} + \varepsilon\bar{\varepsilon} k_n \Psi_m^{(A)},
\end{equation}
is a combined term associated with payoff parameters. Similarly, we apply the same operations to the denominator and finally obtain
\begin{equation}
	= (1 - \alpha) \left( \bar{\varepsilon} k_n + \varepsilon k_n + \bar{\varepsilon} k_m + \varepsilon k_m \right) + \alpha \cdot \Phi_2,
\end{equation}
\begin{equation}
	\Phi_2 = \bar{\varepsilon} k_n \Psi_n^{(A)} + \varepsilon k_n \Psi_n^{(B)} + \bar{\varepsilon} k_m \Psi_m^{(A)} + \varepsilon k_m \Psi_m^{(B)}.
\end{equation}
After combination, Eq. (\ref{13}) can be rearranged as
\begin{equation}
	P_{n \to m} = \frac{(1 - \alpha)\left( \bar{\varepsilon}^2 k_m + \varepsilon^2 k_m + 2\varepsilon\bar{\varepsilon} k_n \right) + \alpha \Phi_1}{(1 - \alpha)\left( \bar{\varepsilon} k_n + \varepsilon k_n + \bar{\varepsilon} k_m + \varepsilon k_m \right) + \alpha \Phi_2},
\end{equation}
For the denominator, we have
$\bar{\varepsilon} k_n + \varepsilon k_n + \bar{\varepsilon} k_m + \varepsilon k_m
=(\bar{\varepsilon}+\varepsilon)(k_n+k_m)
=1\cdot k
=k.$ We therefore conclude that
\begin{equation}\label{22}
	P_{n \to m} = \frac{(1 - \alpha)\left( \bar{\varepsilon}^2 k_m + \varepsilon^2 k_m + 2\varepsilon\bar{\varepsilon} k_n \right) + \alpha \Phi_1}{(1 - \alpha)k + \alpha \Phi_2}.
\end{equation}
Since $\alpha$ is an infinitesimal quantity, the following approximate expansion holds:
\begin{equation}
	\frac{1+\mu\alpha}{1+\nu\alpha} = 1 + (\mu - \nu)\alpha + O(\alpha^2).
\end{equation}
Divide both the numerator and denominator of Equation (\ref{22}) by $(1-\alpha)k$ to rewrite the denominator in the form $1+\nu\alpha$:
\begin{equation}
	P_{n \to m} = \frac{\frac{ \bar{\varepsilon}^2 k_m + \varepsilon^2 k_m  + 2\varepsilon\bar{\varepsilon} k_n }{k} + \frac{\alpha}{(1-\alpha)k} \Phi_1}{1 + \frac{\alpha \Phi_2}{(1-\alpha)k}}.
\end{equation}
Since $\alpha$ is sufficiently small, we have $\frac{1}{1-\alpha}\approx1+O(\alpha)$. As a result, the numerator can be simplified to
\begin{equation}
	\underbrace{\frac{(\bar{\varepsilon}^2 + \varepsilon^2) k_m + 2\varepsilon\bar{\varepsilon} k_n}{k}}_{A}
	+ \underbrace{\frac{\Phi_1}{k}}_{\mu} \cdot \alpha + O(\alpha^2).
\end{equation}
The denominator can be written as
\begin{equation}
	1 + \frac{\alpha \Phi_2}{(1-\alpha)k} \approx 1+\underbrace{\frac{\Phi_2}{k}}_{\nu} \cdot \alpha + O(\alpha^2).
\end{equation}
We thus obtain
\begin{equation}
	\begin{aligned}
		P_{n\to m}&=\frac{A+\mu\alpha+O(\alpha^2)}{1+\nu\alpha+O(\alpha^2)}=\frac{A(1+\frac{\mu}{A}\alpha)}{1+\nu\alpha} + O(\alpha^2)\\
		&= A \left[ 1 + \left( \frac{\Phi_1}{Ak} - \frac{\Phi_2}{k} \right) \alpha \right] + O(\alpha^2)\\
		&=A + \alpha \left( \frac{\Phi_1}{k} - A \cdot \frac{\Phi_2}{k} \right) + O(\alpha^2)\\
		&=\frac{(\bar{\varepsilon}^2 + \varepsilon^2)k_m + 2\varepsilon\bar{\varepsilon}k_n}{k}\\
		&+ \alpha \underbrace{\left[ \frac{\Phi_1}{k} - \frac{(\bar{\varepsilon}^2 + \varepsilon^2)k_m + 2\varepsilon\bar{\varepsilon}k_n}{k} \cdot \frac{\Phi_2}{k} \right]}_{\varpi }
		+ O(\alpha^2).
	\end{aligned}
\end{equation}
Given $k_n=k-k_m$ and $\bar{\varepsilon}=1-\varepsilon$, substitute them into the first term of the above formula and simplify to get
\begin{equation}\label{28}
	P_{n\to m}=2\varepsilon(1-\varepsilon) + \frac{(1-2\varepsilon)^2 k_m}{k} + \alpha \cdot \varpi 
	+ O(\alpha^2).
\end{equation}

\subsection{Population Evolution Equation}
Recall the definitions presented above. Under the assumption of a large-scale network, let $p_m$ denote the proportion of legitimate UAVs adopting the deceptive strategy, and $P_a$ denote the proportion of malicious neighbors implementing the attack strategy. For any legitimate UAV, suppose there are $k_o$ ordinary neighbors among its $k$ neighbors; then the number of malicious neighbors is $k_b=\beta k_o$. Accordingly, the expected total number of neighbors adopting the deceptive strategy (or the attack strategy, both of which manifest as reporting reversed values) is given by:
\begin{equation}
	\mathbb{E}[k_m] = \mathbb{E}[k_{o,m}] + \mathbb{E}[k_{b,m}] = k_o p_m + k_b P_a = k_o (p_m + \beta P_a).
\end{equation}
The expected total number of neighbors satisfies $\mathbb{E}[k]=k_o(1+\beta)$. Therefore,
\begin{equation}\label{30}
	\mathbb{E}\left[ \frac{k_m}{k} \right] = \frac{p_m + \beta P_a}{1 + \beta}.
\end{equation}

Based on the death-birth updating rule, the change rate of the proportion $p_m$ of the deceptive strategy satisfies the following relation \cite{DB}:
\begin{equation}\label{31}
	\hat{p}_m = \mathbb{E}\left[ P_{n\to m} \right] - p_m.
\end{equation}
Substitute Eq. (\ref{28}) into (\ref{31}) and take the expectation over the neighbor distribution to yield
\begin{equation}
	\begin{aligned}
		\hat{p}_m &= \mathbb{E}\left[ 2\varepsilon(1-\varepsilon) + (1-2\varepsilon)^2 \frac{k_m}{k} 
		+ \alpha \cdot\varpi(k_m) \right] \\
		&- p_m + O(\alpha^2).
	\end{aligned}
\end{equation}
Substituting Eq. (\ref{30}) into the above expression, we obtain
\begin{equation}
	\begin{aligned}
		\hat{p}_m &= 2\varepsilon(1-\varepsilon) + (1-2\varepsilon)^2 \frac{p_m + \beta P_a}{1+\beta} - p_m \\
		&+ \alpha \mathbb{E}[\varpi (k_m)] + O(\alpha^2).
	\end{aligned}
\end{equation}

Since the weak selection coefficient $\alpha\ll1$, and $\mathbb{E}[\varpi(k_m)]$ is a bounded constant determined by the model parameters, the term $\alpha\mathbb{E}[\varpi (k_m)]$ constitutes a higher-order infinitesimal. In the limit of $\alpha \to 0$, this term tends to zero, and the dominant contribution to the system evolution comes from the zero-order term. In the subsequent analysis, this paper adopts the zero-order approximation, that is, neglects the term containing $\alpha$ and retains only the $O(1)$ component. Consequently, we obtain the zero-order evolution equation as follows:
\begin{equation}\label{34}
	\begin{aligned}
		&\hat{p}_m \approx  2\varepsilon(1-\varepsilon) + (1-2\varepsilon)^2 \frac{p_m + \beta P_a}{1+\beta} - p_m\\
		&= -p_m \left[ 1 - \frac{(1-2\varepsilon)^2}{1+\beta} \right] + 2\varepsilon(1-\varepsilon) + \frac{(1-2\varepsilon)^2 \beta P_a}{1+\beta}.
	\end{aligned}
\end{equation}
This is a linear ordinary differential equation with respect to $p_m$, which can be rewritten in the standard form of linear ordinary differential equations:
\begin{equation}
	\hat{p}_m=Ap_m+B,
\end{equation}
\begin{equation}
	A=-1+\frac{(1-2\varepsilon)^2}{1+\beta},
\end{equation}
\begin{equation}
	B=2\varepsilon(1-\varepsilon) + \frac{(1-2\varepsilon)^2 \beta P_a}{1+\beta}
\end{equation}
The equilibrium point $p_m^{\text{ESS}}$ is solved from $\dot{p}_m=0$ as $p_m^{\text{ESS}}=-B/A$, whose expanded form is shown below:
\begin{equation}
	p_m^{\text{ESS}} = \frac{2\varepsilon(1-\varepsilon)(1+\beta) + (1-2\varepsilon)^2 \beta P_a}{(1+\beta) - (1-2\varepsilon)^2}.
\end{equation}
Define the error variable $\varrho =p_m-p_m^{\text{ESS}}$. Substituting it into the equation yields
\begin{equation}
	\begin{aligned}
		\frac{d\varrho}{dt}&=\hat{p}_m=Ap_m+B=A(\varrho+p_m^{\text{ESS}} )+B\\
		&=A\varrho+A(-B/A)+B=A\varrho(t)
	\end{aligned}
\end{equation}
The solution to this equation is $\varrho=\varrho(0)e^{At}$. Hence, the stability of the equilibrium point is entirely determined by the sign of coefficient $A$. In practical UAV swarm scenarios, the detection error generally satisfies $0<\varepsilon\le0.5$, and the proportion of malicious UAVs satisfies $\beta>0$. Under such conditions, we always have $A<0$. The exponential term $e^{At}$ decays exponentially over time such that $\varrho(t)\to0$. The error gradually diminishes and eventually converges back to the equilibrium point, meaning the system is stable. Therefore, the equilibrium $p_m^{\text{ESS}}$ under the zero-order approximation is asymptotically stable.

\begin{figure}[t]
	\centerline{\includegraphics[width=3.3in,keepaspectratio]{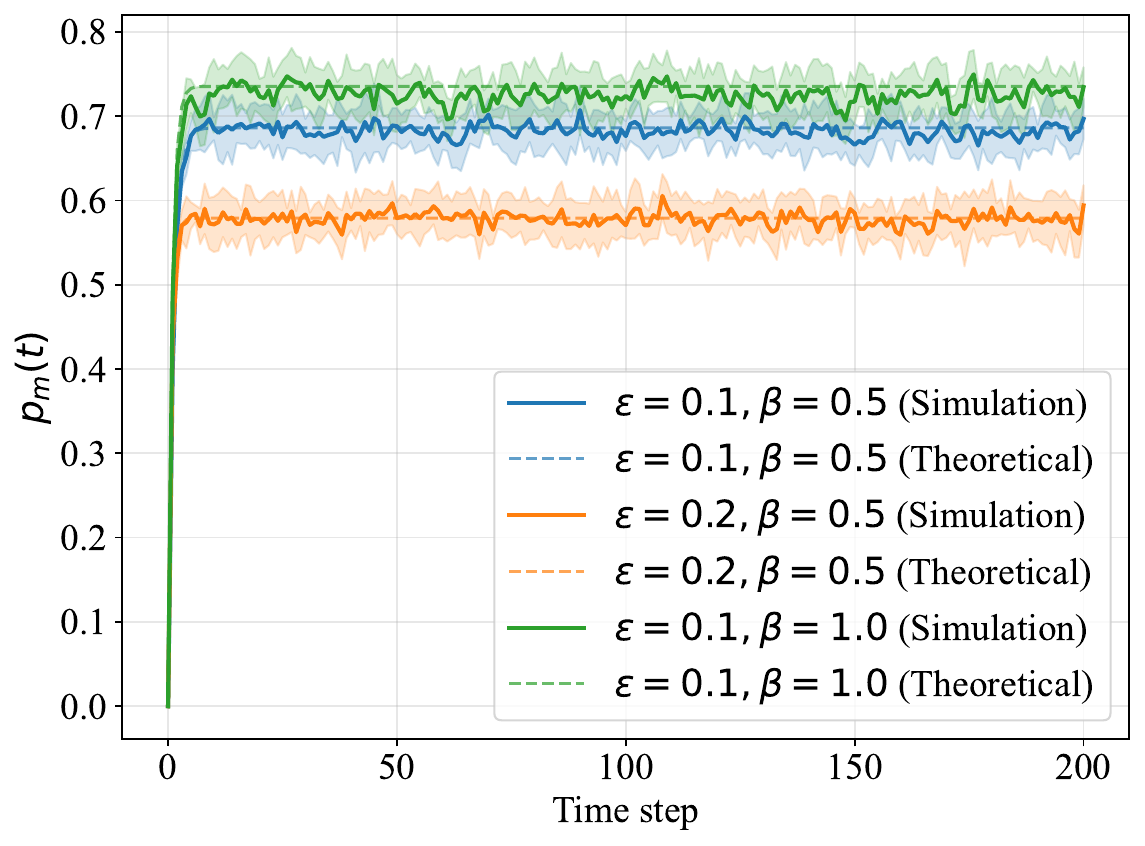}}
	\caption{Evolution under different parameters (regular network), $P_a=1.0$}
	\label{fig1}
\end{figure}

\section{Simulation Experiments and Analysis}\label{Sec4}
In this section, the Monte Carlo simulation method is adopted to systematically verify the evolutionary dynamics equations, evolutionarily stable states and parameter influences. The number of Monte Carlo trials is set to 10. Regular networks are mainly utilized for simulations, where each UAV possesses the same number of neighbor nodes with a fixed degree $k=10$. In addition, to verify the robustness of the proposed model against network topologies, BA scale-free networks and ER random networks are adopted in partial experiments. The number of UAVs in all three types of networks is $N=500$. The average degree of the BA network is approximately 6, and the connection probability of the ER network is set to $p=k/N=0.02$ to maintain an average degree of about 10. The payoff matrix parameters are set as $u_{ns}=0.8$, $u_{nd}=0.6$, $u_{ms}=0.6$, and $u_{md}=0.4$. The initial proportion of deceptive strategies is initialized to 0. The evolution time is fixed to 200 to reach a steady state.

In Fig. \ref{fig1}, the regular network is adopted as the fixed network topology to compare the evolutionary processes under three groups of parameter settings. For the set $\varepsilon=0.1, \beta=0.5$, the proportion rises rapidly starting from zero and eventually converges to a steady value. When the observation error is increased to $\varepsilon=0.2$ with all other parameters unchanged, the steady-state value drops below 0.6. This phenomenon indicates that as the observation error increases, UAVs hold lower confidence in their own observations, and their decision-making behaviors become closer to random choices, which to some extent weakens the directional induction effect of malicious nodes. When the proportion of malicious nodes is raised to $\beta=1.0$, the steady-state value rises above 0.7. This demonstrates that an increase in the number of malicious nodes can significantly amplify the attack effect.

\begin{figure}[t]
	\centerline{\includegraphics[width=3.3in,keepaspectratio]{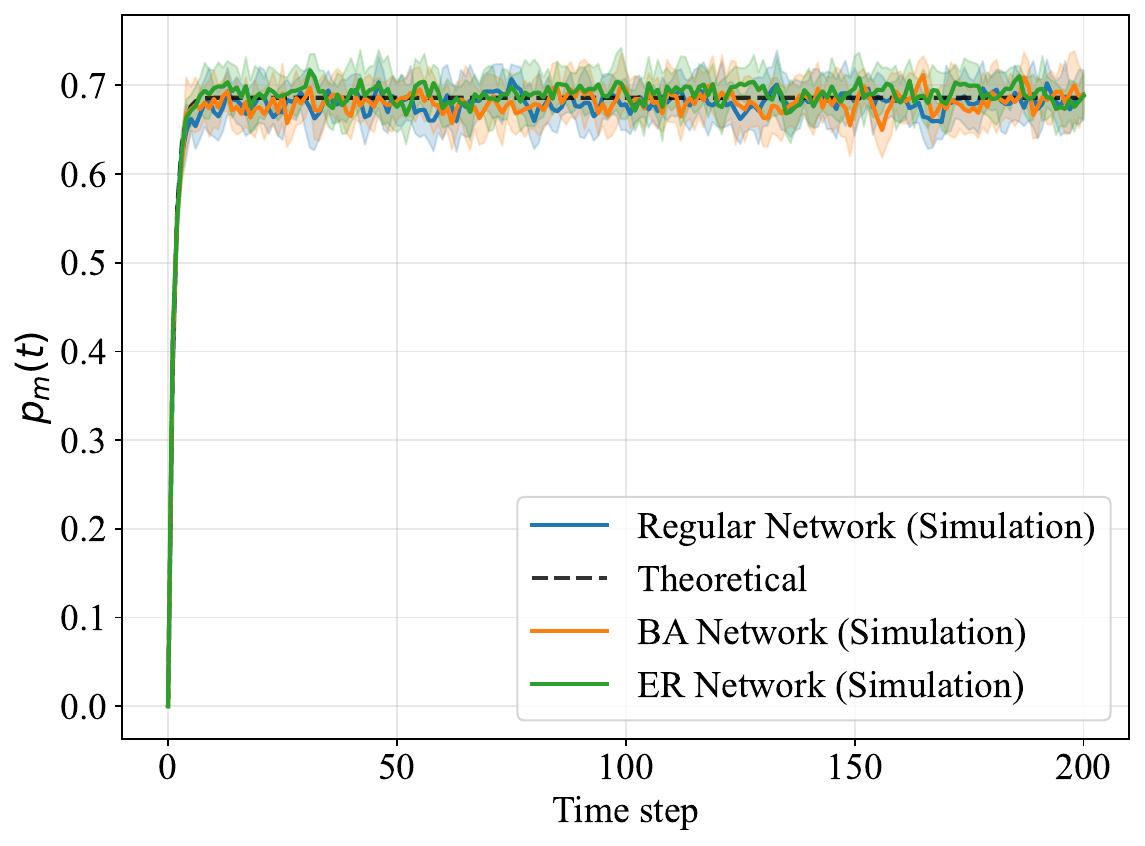}}
	\caption{Different network topologies evolution, $\varepsilon=0.1, \beta=0.5, P_a=1.0$}
	\label{fig2}
\end{figure}

In Fig. \ref{fig2}, the parameters are fixed as $\varepsilon=0.1, \beta=0.5, P_a=1.0$, and the evolutionary processes under three distinct network topologies are compared. The three evolutionary curves almost completely overlap, and all match well with the theoretical prediction values represented by the black dashed line. This result verifies the robustness of the proposed model. Specifically, the network topology exerts little influence on the macroscopic evolutionary trend of the deceptive strategy, which proves that the theoretical model put forward in this paper is applicable to various practical network scenarios.

\begin{figure}[t]
	\centerline{\includegraphics[width=3.3in,keepaspectratio]{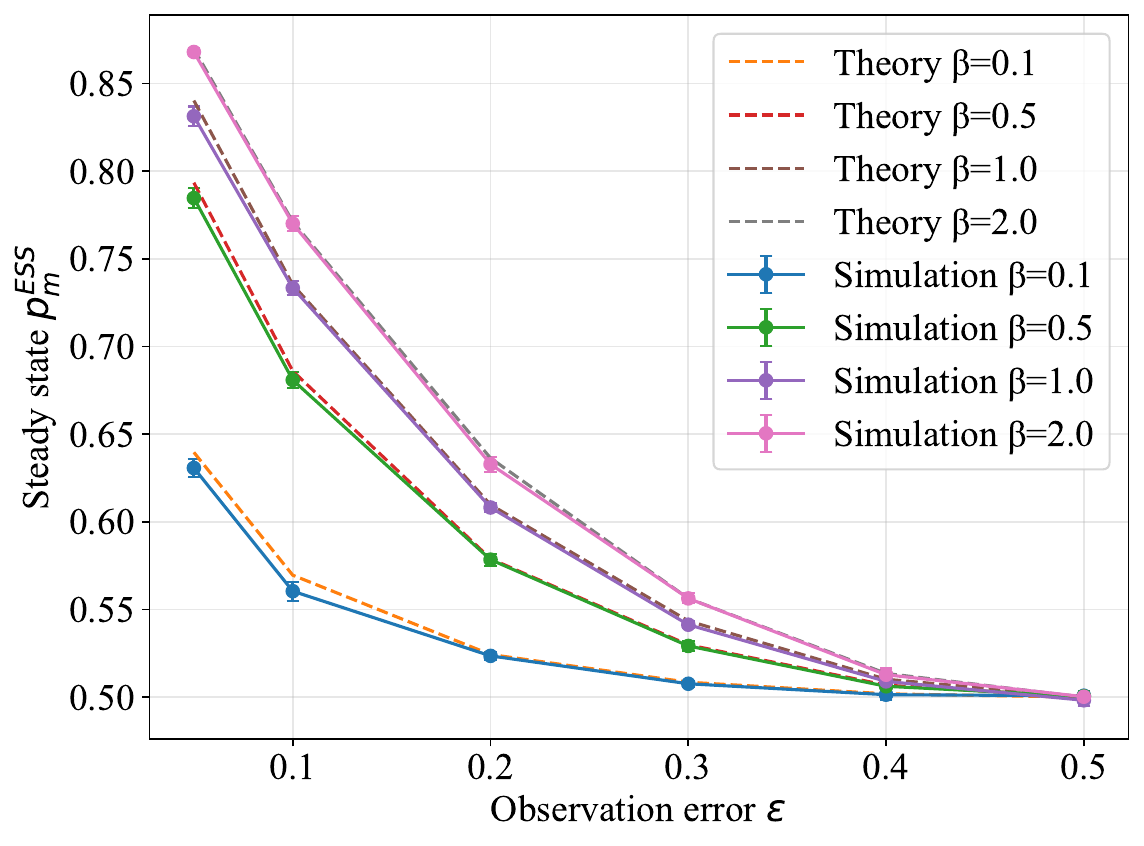}}
	\caption{Observation error $\varepsilon$ vs $p^{ESS}_m$ under different $\beta$, $P_a=1$}
	\label{fig3}
\end{figure}
\begin{figure}[t]
	\centerline{\includegraphics[width=3.3in,keepaspectratio]{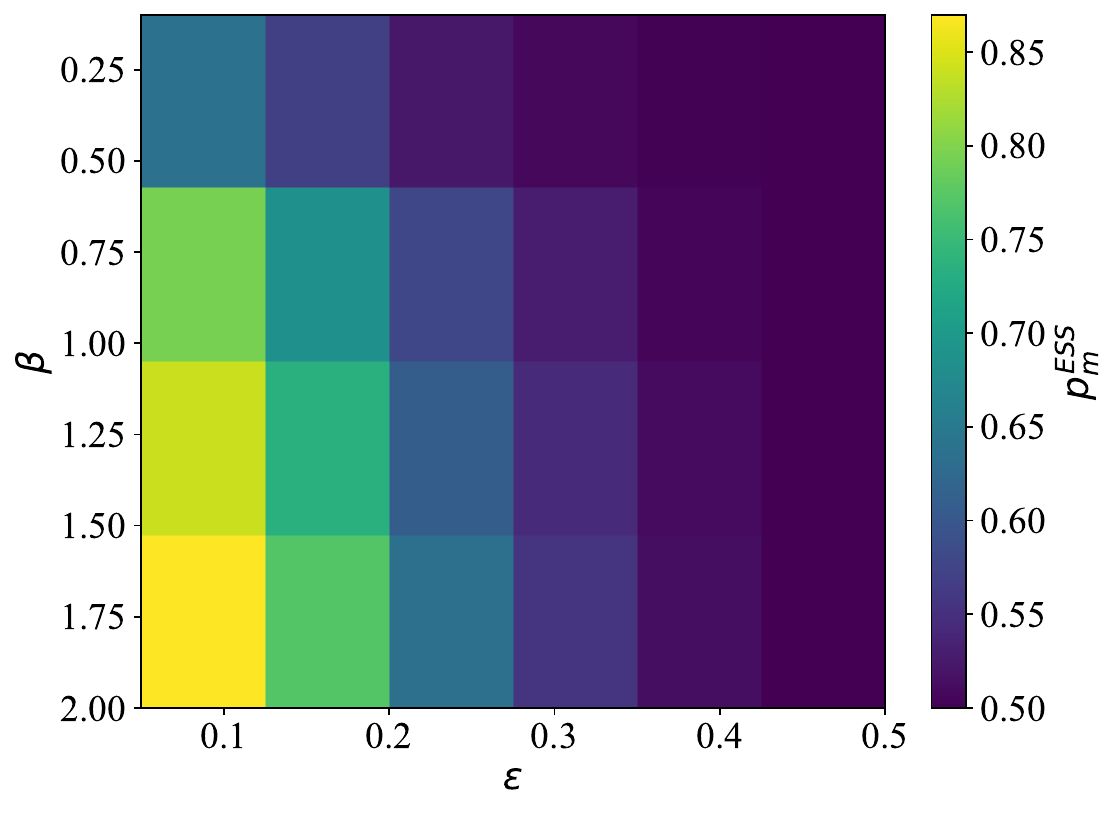}}
	\caption{Theoretical ESS Heatmap}
	\label{fig4}
\end{figure}

The results shown in Figs. \ref{fig3} and \ref{fig4} reveal that the observation error $\varepsilon$ and the proportion of malicious nodes $\beta$ are two key factors affecting the evolutionarily stable state of the system, yet they exert opposite effects. The equilibrium proportion $p_m^{\text{ESS}}$ decreases monotonically as $\varepsilon$ grows. The underlying reason is that when the observation error is small, legitimate UAVs can perceive the true system state relatively accurately. In this case, the deceptive strategies induced by malicious nodes can spread effectively, keeping the steady-state proportion of deceptive strategies at a high level. As the observation error increases, the directional induction effect of malicious nodes is diluted, as analyzed in Fig. \ref{fig1}. In contrast, $p_m^{\text{ESS}}$ rises with the growth of the malicious node proportion $\beta$. This implies that a higher proportion of malicious nodes in the network raises the probability for legitimate UAVs to have more malicious neighbors around them, making them more vulnerable to deception and inclined to adopt erroneous strategies. When $\varepsilon=0.5$, the steady-state values of all curves converge to 0.5. This is because the observation results become completely random, and UAVs are unable to extract any valid information from their observations.

\begin{figure}[t]
	\centerline{\includegraphics[width=3.3in,keepaspectratio]{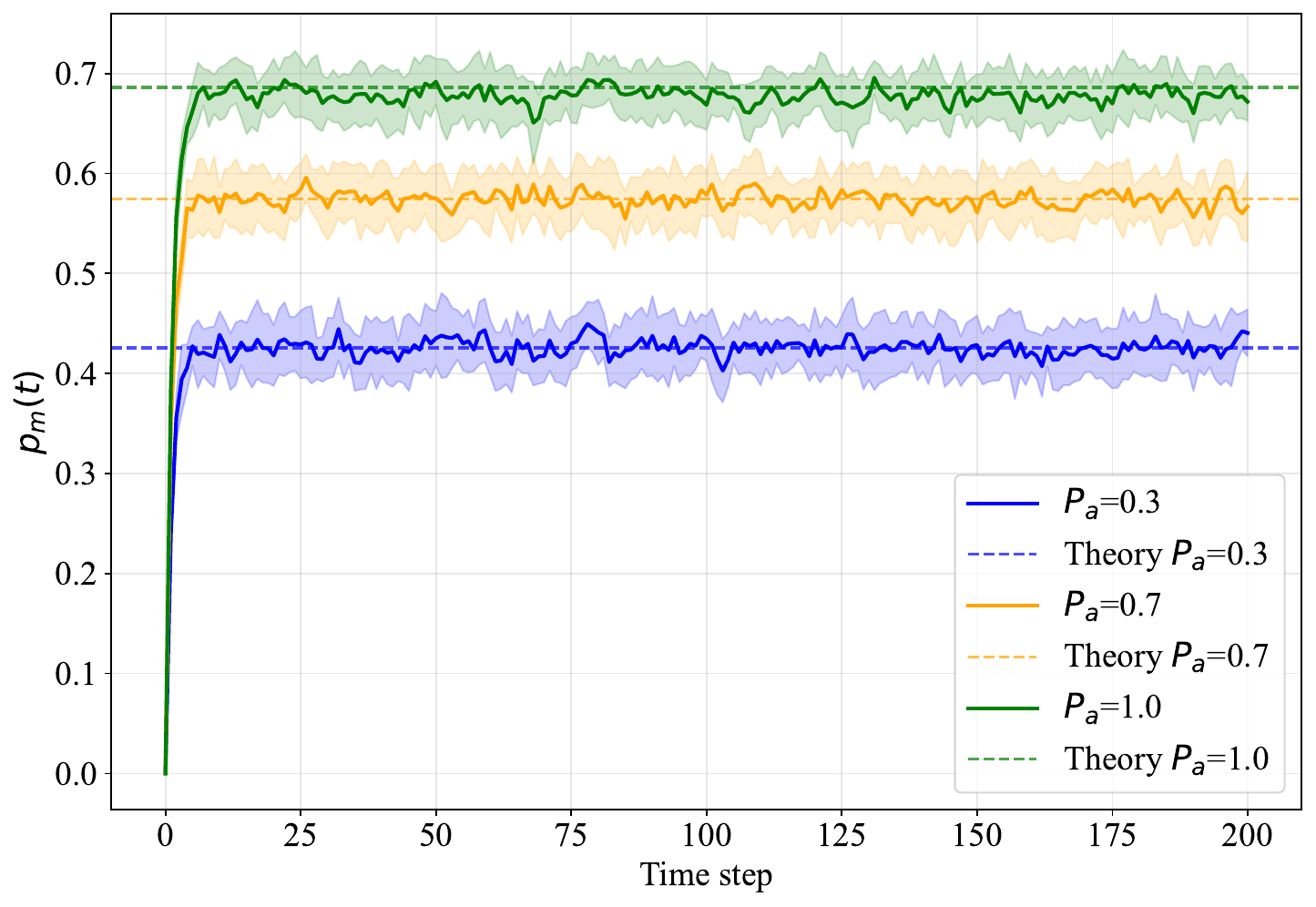}}
	\caption{Impact of Attack Probability $P_a$, $\varepsilon=0.1, \beta=0.5$.}
	\label{fig5}
\end{figure}

Fig. \ref{fig5} plots the evolutionary curves of $p_m(t)$ under different values of $P_a$. The steady-state value reaches its maximum when $P_a=1.0$, followed by the case of $P_a=0.7$, while the steady-state value is the lowest at a smaller $P_a$. This further illustrates that a stronger attack intensity of malicious nodes produces a more prominent interference effect on the decision-making of legitimate UAVs, inducing more normal UAVs to adopt deceptive strategies.

\section{Conclusion}\label{Sec5}
Aiming at UAV swarm under Byzantine attack, this paper took local conformity behaviors into consideration and established an evolutionary model based on graph evolutionary game theory. With the death-birth updating rule, we derived the macroscopic dynamic equation of the proportion of deceptive strategies as well as the analytical expression of the ESS. Through theoretical analysis and simulation verification, it was concluded that the conformity effect acts as an amplifier for Byzantine attacks. Malicious nodes not only disseminated false information on their own, but also induced legitimate UAV nodes to follow suit, thereby magnifying their destructive power within the swarm. This characteristic is not reflected in traditional Byzantine models. This study provides a theoretical reference for designing security strategies against Byzantine attacks in UAV swarm.

\end{document}